\documentclass[aps,preprint,superscriptaddress,onecolumn,floatfix]{revtex4}
\usepackage{latexsym}
\usepackage{amsfonts}
\usepackage{amssymb}
\usepackage{amsmath}
\usepackage{graphicx}

\begin{document}

\title{Large microwave generation from d.c. driven magnetic vortex oscillators in magnetic tunnel junctions}

\author{A. Dussaux}
\affiliation{Unit\'e Mixte de Physique CNRS/Thales and Universit\'e Paris Sud 11, 1 ave A. Fresnel, 91767 Palaiseau, France}
\author{B. Georges}
\affiliation{Unit\'e Mixte de Physique CNRS/Thales and Universit\'e Paris Sud 11, 1 ave A. Fresnel, 91767 Palaiseau, France}
\author{J. Grollier}
\affiliation{Unit\'e Mixte de Physique CNRS/Thales and Universit\'e Paris Sud 11, 1 ave A. Fresnel, 91767 Palaiseau, France}
\author{V. Cros}
\affiliation{Unit\'e Mixte de Physique CNRS/Thales and Universit\'e Paris Sud 11, 1 ave A. Fresnel, 91767 Palaiseau, France}
\author{A. V. Khvalkovskiy}
\affiliation{Unit\'e Mixte de Physique CNRS/Thales and Universit\'e Paris Sud 11, 1 ave A. Fresnel, 91767 Palaiseau, France}
\affiliation{A.M. Prokhorov General Physics Institute of RAS, Vavilova str. 38, 119991 Moscow, Russia}
\author{A. Fukushima}
\affiliation{National Institute of Advanced Industrial Science and Technology (AIST) 1-1-1 Umezono, Tsukuba, Ibaraki 305-8568, Japan}
\author{M. Konoto}
\affiliation{National Institute of Advanced Industrial Science and Technology (AIST) 1-1-1 Umezono, Tsukuba, Ibaraki 305-8568, Japan}
\author{H. Kubota}
\affiliation{National Institute of Advanced Industrial Science and Technology (AIST) 1-1-1 Umezono, Tsukuba, Ibaraki 305-8568, Japan}
\author{K. Yakushiji}
\affiliation{National Institute of Advanced Industrial Science and Technology (AIST) 1-1-1 Umezono, Tsukuba, Ibaraki 305-8568, Japan}
\author{S. Yuasa}
\affiliation{National Institute of Advanced Industrial Science and Technology (AIST) 1-1-1 Umezono, Tsukuba, Ibaraki 305-8568, Japan}
\author{K.A. Zvezdin}
\affiliation{A.M. Prokhorov General Physics Institute of RAS, Vavilova str. 38, 119991 Moscow, Russia}
\affiliation{Istituto P.M. s.r.l., via Cernaia 24, 10122 Torino, Italy}
\author{K. Ando}
\affiliation{National Institute of Advanced Industrial Science and Technology (AIST) 1-1-1 Umezono, Tsukuba, Ibaraki 305-8568, Japan}
\author{A. Fert}
\affiliation{Unit\'e Mixte de Physique CNRS/Thales and Universit\'e Paris Sud 11, 1 ave A. Fresnel, 91767 Palaiseau, France}

\begin{abstract}
Spin polarized current can excite the magnetization of a ferromagnet through the transfer of spin angular momentum to the local spin system. This pure spin-related transport phenomena leads to alluring possibilities for the achievement of a nanometer scale, CMOS compatible and tunable microwave generator operating at low bias for future wireless communications. Microwave emission generated by the persitent motion of magnetic vortices induced by spin transfer effect seems to be a unique manner to reach appropriate spectral linewidth. However, in metallic systems, where such vortex oscillations have been observed, the resulting microwave power is much too small. Here we present experimental evidences of spin-transfer induced core vortex precessions in MgO-based magnetic tunnel junctions with similar good spectral quality but an emitted power at least one order of magnitude stronger. More importantly, unlike to others spin transfer excitations, the thorough comparison between experimental results and models provide a clear textbook illustration of the mechanisms of vortex precessions induced by spin transfer. 
\end{abstract}

\maketitle

The spin transfer torque due to the interaction between a spin polarized current and a magnetization \cite{Slonczewski:JMMM:1996,Berger:PRB:1996} has led in the last decade to the emergence of a number of interesting novel effects such as current induced magnetization reversal \cite{Katine:PRL:2000, Grollier:APL:2001} or self-sustained magnetization oscillations \cite{Kiselev:Nature:2003,Rippard:PRL:2004}. The observation of spin transfer induced magnetization precession brings some promising possibilities for designing a new type of nanoscale microwave oscillators, the so-called spin transfer nano oscillators (STNOs), capable to compete in terms of high frequency caracteristics with currently used millimeter-scale microwave synthetizer. One of the tantalizing of the STNOs is that they are tunable over a wide frequency range by varying the applied dc current or magnetic field. In most experiments up to now \cite{Kiselev:Nature:2003,Rippard:PRL:2004}, the current-induced excitations were  \textit{quasi uniform} precessions in purely metallic structures, which leads to a much too small generation of microwave power for the majority of applications. Various solutions have been recently proposed, for example by synchronizing an assembly of STNOs \cite{Kaka:Nature:2005,Mancoff:Nature:2005,Ruotolo:NatNanotech:2009,Georges:APL:2008,Georges:PRL:2008} or by using MgO based Magnetic Tunnel Junctions (MTJs) \cite{Yuasa:NatMat:2004,Parkin:NatMat:2004} that deliver much larger power because of the larger magnetoresistance \cite{Deac:NatPhys:2008,Nazarov:APL:2006,Houssamedine:APL:2008}. However the spectral linewidths with MgO MTJs are still very large ($\approx$ 100 MHz) for standard excitations of the free layer in MTJs due to the chaotization of the magnetic system induced by the spin transfer torque and/or the high non-linearity \cite{Georges:arxiv:2009}. 

An alternative approach using the current driven motion of a magnetic vortex as source of microwave power has been recently studied in metallic nanopillars \cite{Pribiag:NatPhys:2007}. A very interesting mode is the circular motion of the vortex core around its equilibrium position \cite{Guslienko:JAP:2002}. This low energy mode gives rise to very small linewidths (less than 1 MHz) in the sub-Gigahertz range but generates only a very small integrated power in all metallic devices (about 0.01 nW). In the present paper, we show that large microwave powers and narrow linewidths can be obtained simultaneously by spin transfer induced vortex motion in MgO MTJs. 

The studied devices are circular shape nanopillars with a diameter $d$ = 170 nm patterned from the whole magnetic stack : synthetic antiferromagnet / MgO 1.075 nm/ NiFe 15 nm/capping. For the chosen parameters of NiFe thickness and dot diameter, we have checked that the most favorable configuration is a single vortex state (see Methods). 
In these nano-MTJs, the vortex dynamics is converted into a microwave signal thanks to the large magnetoresistive ratio of the Tunnel Magnetoresistance effect (TMR). This enable us to detect not only large sustained vortex oscillations induced by spin transfer but also low signal associated to the ferromagnetic resonance of the thermally excited vortex. Such resonant motion is efficient enough to produce competitive passive microwave elements such as high gain resonators \cite{Nozaki:APL:2009,Kasai:APEX:2009}.

\begin{figure}
   \centering
    \includegraphics[width=12 cm]{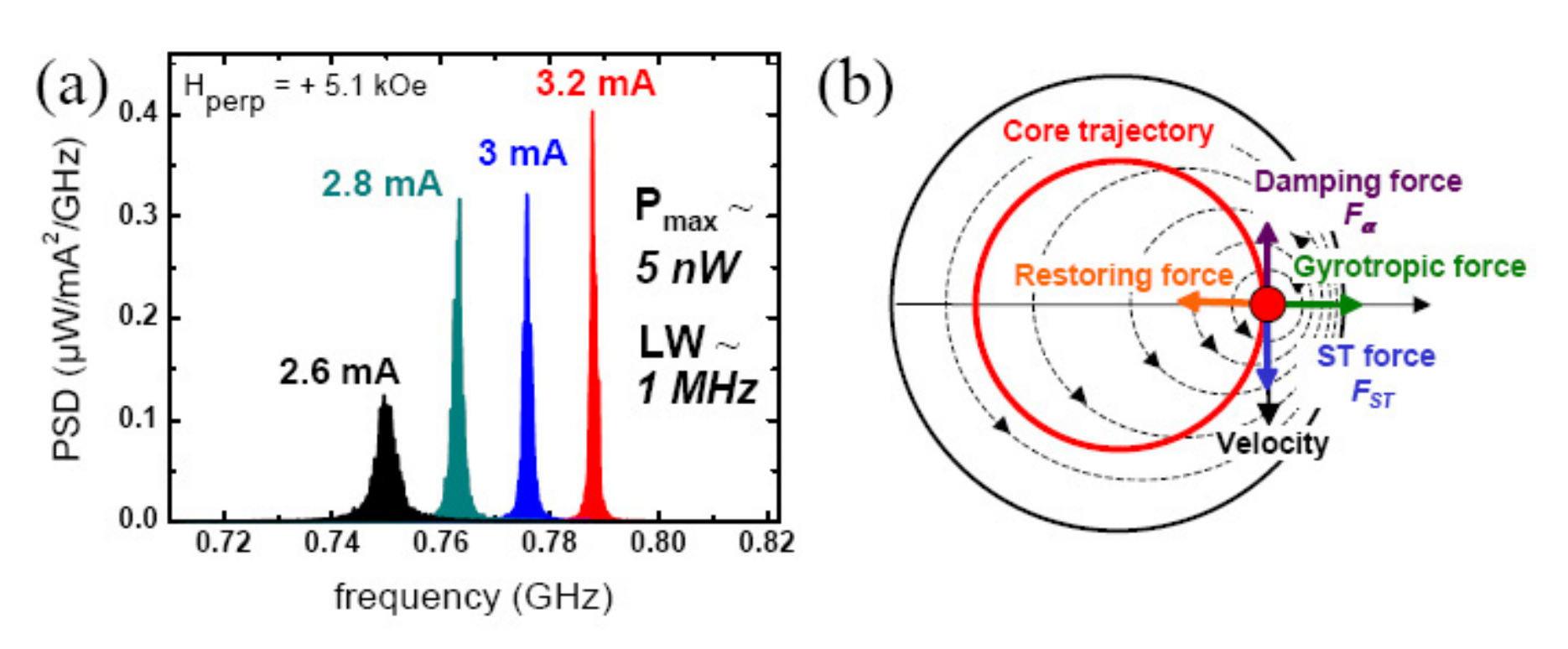}
     \caption{(a) Power spectral densities (PSD) normalized by $I_{dc}^{2}$, obtained for $I_{dc}$ = 2.6, 2.8, 3 and 3.2 mA with  $H_{perp}$ = 5.1 kOe. (b) Schematic of the different forces acting on the vortex core.}
\label{fig1}
\end{figure}

In Fig.\ref{fig1} (a), we display some microwave emission spectra recorded with $H_{perp}$ = + 5.1 kOe  at several positive current values. For all $I_{dc}$, a single peak is observed in the sub-GHz frequency range that is characteristic of the coherent gyrotropic motion of the magnetic vortex core induced by the spin transfer torque \cite{Pribiag:NatPhys:2007}. The emission frequency increases with $I_{dc}$ (about 80 MHz/mA) together with a large increase of the peak power spectral density (PSD). The best results have been obtained at large currents ($I_{dc}$= 3.2 mA) but still below the threshold above which the MTJ is damaged. At this value, the peak linewidth is only 1.1 MHz and the integrated power is about 5 nW. This is to our knowledge the highest emitted power for STNOs showing similar linewidth. It therefore demonstrates the interest to use the motion of a vortex as microwave source to benefit from the advantages of both the coherency of the vortex gyrotropic motion and the large magnetoresistive signal provided by a MTJ.

In most of cases, spin transfer excitations that is focused on the dynamics of a \textit{quasi uniform} magnetization, is suffering from a lack of comparison with theoretical predictions. The reason comes mainly from the difficulties to accurately take into account the presence of multi-modes, the different sources of spin transfer torques, the role of the Oersted field or of the temperature \cite{Stiles:spin:2006}. An important impact of our work is that we provide a definite comparison and an excellent agreement between experimental data, micromagnetic simulations and theoretical predictions. As a matter of fact, we will show that our experiments can be interpreted in the frame of the most recent theoretical descriptions of the spin-transfer induced vortex gyrotropic motion using a modified Thiele equation \cite{Khvalkovskiy:arxiv:2009,Guslienko:arxiv:2010}. The starting point is the modified Thiele equation taking into account the spin transfer torque :

\begin{equation}\label{ThieleEquation}
\mathbf{G}\times\frac{d\mathbf{X}}{dt} - \frac{\partial W}{\partial \mathbf{X}} - \hat{D} \frac{d\mathbf{X}}{dt} + \mathbf{F}_{ST}=0, 
\end{equation}

where $\textbf{G}= -2 \pi p L M_s / \gamma \textbf{$e_z$}$ is the gyrovector, $W(\textbf{X})$ is the potential energy of the off-centered vortex, $\hat{D}$ is the damping dyadic. 
In Fig.\ref{fig1} (b), we depict the four forces responsible for the rotational motion of the vortex core around its equilibrium position. In Eq.(\ref{ThieleEquation}), the first two terms, corresponding respectively to the gyrotropic force and the restoring force, are radial with respect to the vortex trajectory and their balance sets the frequency of the motion. The last two terms, corresponding respectively to the viscous damping (that latter on, we call $F_{\alpha}$) and the spin transfer force $F_{ST}$ are tangent to the trajectory, therefore their compensation sets the amplitude of the orbit. Note that we introduce only the Slonczewski term for the spin transfer torque since the Field like term is equivalent to an additional applied magnetic field directed along the polarizer and its amplitude is negligeable compared to $H_{perp}$ we apply \cite{Oh:NatPhys:2009} . In the case of a fixed and uniform polarizer, we have derived an expression for the spin transfer force $F_{ST}$ given by \cite{Khvalkovskiy:arxiv:2009} :

\begin{equation}\label{STT}
F_{ST}=  \sigma \pi J p_z M_s L a   \vec{e}_\chi,
\end{equation}

where $\sigma$ is the spin transfer efficiency equal to $\hbar P_{spin}/2 \left|e\right| L M_{s}$ with $P_{spin}$ the spin polarization of the current, L the thickness of the dot, $J$ the current density (defined as positive for electrons flowing from the NiFe layer to SAF layers), $a$ the orbit radius and $\vec{e}_{\chi}$ the unit vector tangential to the orbit of the core. The spin transfer force $F_{ST}$ is proportional to $p_z$, the out-of-plane component of the polarizer ($p_z$ = $cos \theta$ if $\theta$ is the angle between the magnetization and the out-of-plane direction). The expression of the damping force $F_{\alpha}$ is :
\begin{equation}\label{damping}
F_{\alpha}= - 2 \pi \alpha \eta \frac{M_s}{\gamma} \omega L a p \vec{e}_\chi, 
\end{equation}
where $\alpha$ is the damping constant, $\eta=\hat{D}/ \alpha G$ is a damping parameter, $M_s$ the saturation magnetization of the NiFe layer, $\gamma$ the gyromagnetic ratio and $\omega$ the rotational speed. The sign of the core velocity, and consequently the sign of $F_{\alpha}$ depends on the polarity $p$ of the vortex core that can take two values : $p$ = 1 for the core pointing up and $p$ = -1 when it points down. 

The condition for sustained gyration of the vortex core is that the spin transfer force counterbalances the damping. It follows from the respective signs of equations \ref{STT} and  \ref{damping} that spin-transfer induced vortex oscillations can be observed only if the vortex core and the polarizer are pointing in the same direction when the current is positive, and if they point in opposite directions when the current is negative. Our MgO based nanojunctions with a vortex configuration are model systems compared to metallic devices to test these theoretical predictions. The reasons are twofold. First the SAF structure of the polarizer provides a uniform and fixed spin polarization. Second the current amplitude is small enough not to bend the polarizer through the Oersted field or induce some dynamics into the SAF.

\begin{figure}
   \centering
    \includegraphics[width=8.5 cm]{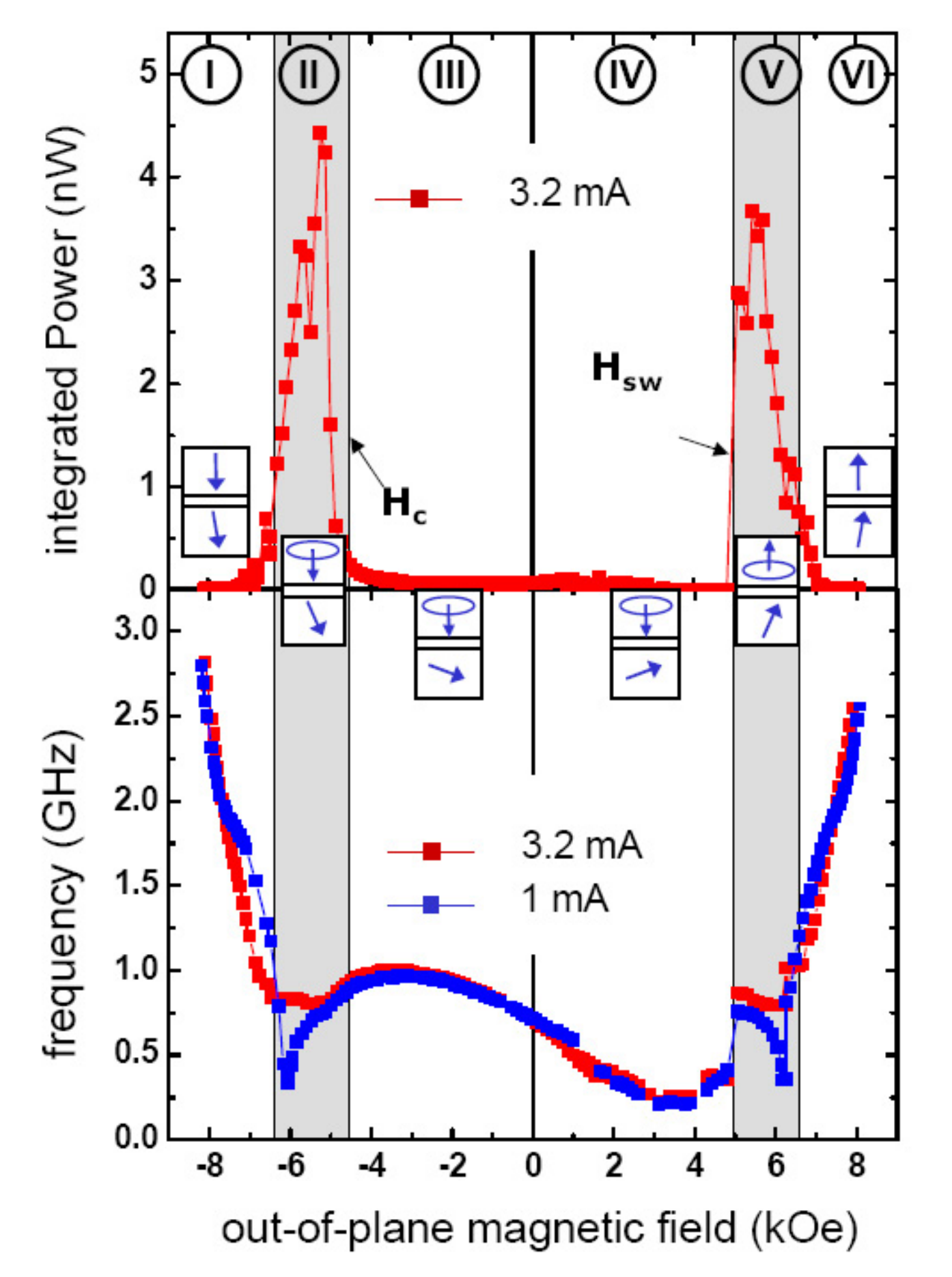}
     \caption{(top) integrated power as a function of the out-of-plane magnetic field. The field is swept from negative to positive values. $H_c$ is the critical field at which microwave oscillations occur for $I_{dc}$ = 3.2 mA, $H_{sw}$ is the field at which the vortex core reverses. (bottom) frequency as a function of the out-of-plane magnetic field. Blue dots : $I_{dc}$ = 1 mA, red dots : $I_{dc}$ = 3.2 mA.  }
\label{fig2}
\end{figure}

To show that, we plot in Fig.\ref{fig2} the integrated microwave power for $I_{dc}$ = 3.2 mA (top panel)and the frequency for two values of $I_{dc}$    (bottom panel) as a function of the out-of-plane magnetic field $H_{perp}$. The field is swept from -8.1 to 8.1 kOe and the out-of-plane component of the polarization $p_z$ is expected to be proportional to $\left|H_{perp}\right|$.

Considering first the variation of the integrated power as a function of the out of plane field in the top panel of Fig.\ref{fig2}, we can identify the following successive zones. In zone I, for $H_{perp}$ $<$ - 7.5 kOe, the absolute value of field is large enough to induce a uniform magnetic configuration (no vortex) and the power is negligibly small. In zone II, a vortex turns out and, as the vortex core polarity and the out-of-plane component of the polarization $H_{z}$ are both negative, a vortex gyration is induced and a large microwave power is emitted (about 5 nW at $I_{dc}$ = 3.2 mA). When $\left|H_{perp}\right|$ becomes smaller than $H_{c}$ $\approx$ 4.5 kOe (zone III), the out of plane polarization $p_z$ is too small to excite large gyrations and the power is negligible at the scale of the figure. In zone IV, not only $\left|p_z\right|$ is too small for $H_{perp}$ $<$ $H_{c}$ but, with the inversion of  $H_{perp}$, the vortex core polarity and the polarizer have opposite signs. The condition for vortex gyration is not satisfied and, even for $H_{perp}$ $>$ $H_{c}$, the power remains very small (max. 270 pW at $I_{dc}$ = 3.2 mA). In zone V that begins at $H_{sw}$ $=$ 5 kOe, the switching of the vortex core aligns $p$ and $p_z$, which induces gyrations and large microwave power. Finally, in zone VI, for  $H_{perp}$ $>$ 6.5 kOe, as in zone I, there is no vortex and the power is negligible. 

\begin{figure}
   \centering
    \includegraphics[width=8.5 cm]{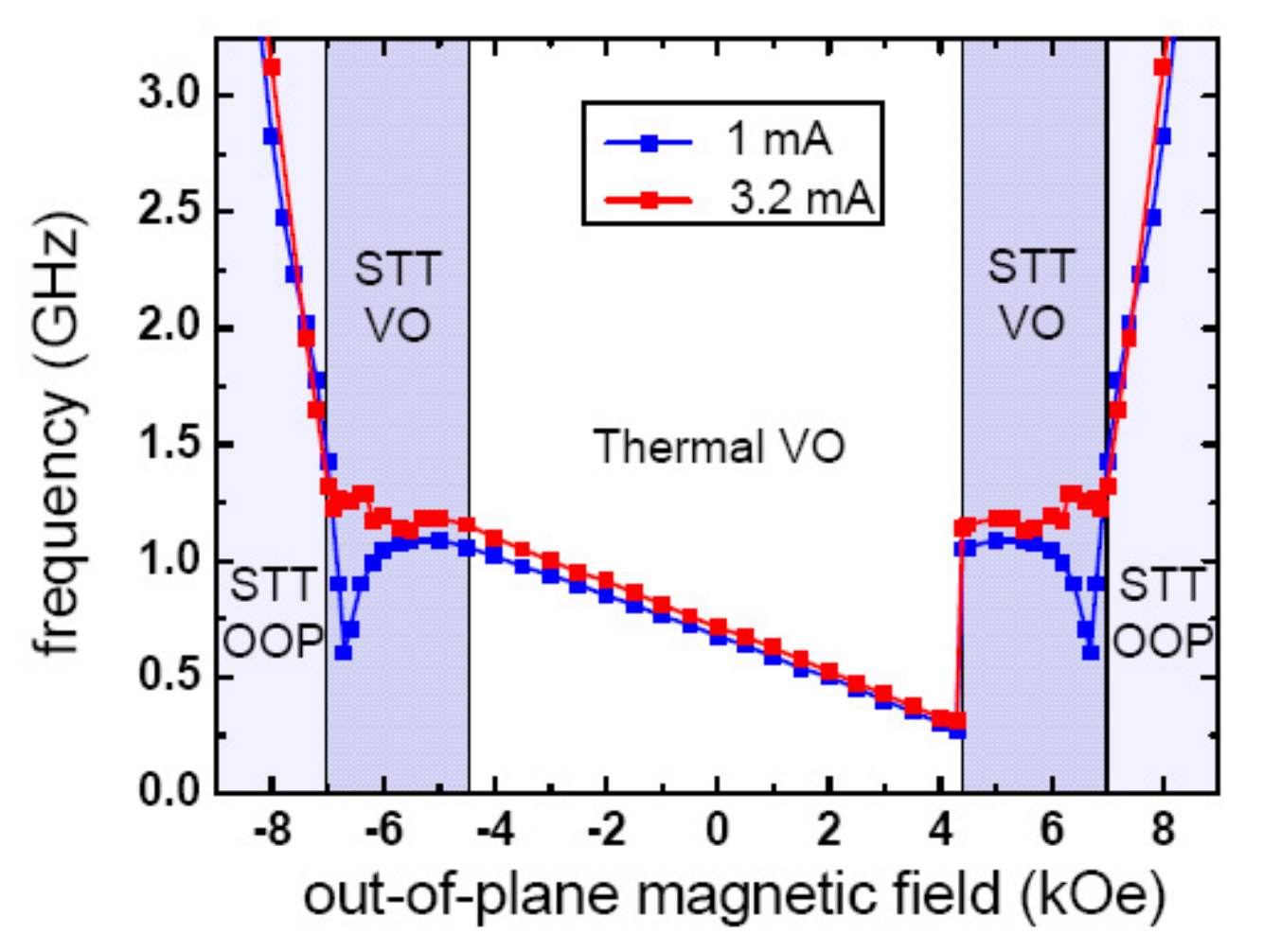}
     \caption{micromagnetic simulations of the frequency versus the out-of-plane magnetic field for $I_{dc}$ = 1 and 3.2 mA. Different regimes are identified i.e out-of-plane uniform precession (STT-OOP), spin transfer vortex oscillation (STT-VO) and thermally excited vortex resonance (Thermal VO).}
\label{fig3}
\end{figure}

The variation of the frequency as a function of the field in the bottom panel of Fig.\ref{fig2} fits with the variation of the power discussed above. 
For $\left|H_{perp}\right|$ $>$ 7.5 kOe (zones I and VI), the magnetic field is strong enough to saturate the magnetization of the NiFe layer for all $I_{dc}$ and a spatially uniform configuration with the magnetization pointing out-of-plane is stabilized. The frequency of the excitations increases linearly with the field independently on the current $I_{dc}$, as expected for small amplitude uniform precessions from the FMR Kittel formula. In this large field region, the slope is close to the expected value of 2.8 GHz/kOe and the intercept with the zero-frequency axis occurs at 7.7 kOe, which corresponds to $M_s^{\small{NiFe}}$ = 9 kOe after taking into account the demagnetizing factors. For $\left|H_{perp}\right|$ between 6.5 and 7.5 kOe, one clearly sees in Fig.\ref{fig2} (a) a slight deviation from the Kittel mode, reflecting the onset of a weak inhomogeneity that is suppressed when $I_{dc}$ increases. Such behavior is well reproduced by the micromagnetic simulations of the frequency versus out-of-plane magnetic field presented in Fig.\ref{fig3} by introducing a small tilt (1$^{\circ}$) of $H_{perp}$ from the normal direction. In both zone I and VI, as mentioned before, very low power emissions are measured in agreement with what is generally observed for uniform magnetization oscillations of the free layer.

The transition between the uniform magnetization and the vortex configurations (zone II) occurs at $H_{perp}$ = - 6.1 kOe (resp. transition from vortex to uniform configuration, zone V, at $H_{perp}$ = 6.1 kOe). It is characterized by a dip in frequency as can be seen from the curve at $I_{dc}$ = 1 mA in Fig.\ref{fig2}. This dip is well reproduced by the micromagnetic simulations shown in Fig.\ref{fig3}, and is related to a modification of the shape of the vortex core that results into a softening of the mode. In this field range of large emission, the simulations for $I_{dc}$ = 3.2 mA are in good agreement with the experimental results. We find that the combined action of the spin transfer torque and the Oersted field strongly modifies the evolution of the frequency and maintains the vortex present at larger $H_{perp}$ (see additional materials for the observation of the gradual disappearance of the frequency dip as the current is increased). In the experiment and simulations, we see that several large-amplitude non-uniform oscillating modes are successively excited when changing the field. Note that such sequential transition of modes is not detectable by measuring the device resistance versus field. Below the critical field $\left|H_{c}\right|$, only thermally excited vortex motion associated with a very small power is observed and in consequence the two frequency curves at different $I_{dc}$ are superimposed. While increasing $H_{perp}$ from - 4.5 kOe to + 4.9 k0e, i.e. zones III and IV, after a small increase of the frequency, the general trend is a decrease of the gyrotropic frequency (see Fig.\ref{fig2}, top panel) as already observed in vortex resonance induced by ac field by De Loubens \textit{et al} (see Eq.2 in Ref.\cite{DeLoubens:PRL:2009}) due to the field dependence of the vortex stiffness. The deviation from the expected linear behavior of the frequency vs. field is attributed to a deformation of the energy landscape by grains. For $H_{sw}$ = + 5 kOe, a large jump in frequency occurs due to the reversal of the vortex core polarity. After the switching, it becomes aligned to the applied field and to the polarizer like in the case of large negative fields (zone I). The reversal of the core polarity, associated with the abrupt change in frequency and the recovery of the oscillations, is well reproduced in the simulation (see Fig.\ref{fig3}). It is worth emphasizing that a symmetrical behavior is obtained when saturating first at large positive $H_{perp}$ and then decreasing the field toward negative values.

As already mentioned, for negative currents, the condition to get large oscillations is that the vortex core polarity and the out-of-plane component have opposite signs. The measurement of the integrated microwave power for $I_{dc}$ = -5 mA as a function of $H_{perp}$, swept from -8.1 to 8.1 kOe is presented as additional material. Large microwave power is detected in a very narrow field window between $H_{c}$ =  3.7 kOe and $H_{sw}$ =  4.3 kOe.  
We notice that $H_{sw}$ is reduced in the case of negative currents because the vortex core velocity is enhanced by spin transfer \cite{Khval:Xarchiv:2009}. It implies that we have to increase the injected absolute value of the current in order to decrease $H_{c}$ and fulfill the condition $H_{c}$  $<$ $H_{sw}$. The counterpart of this additional experimental verification of the model, is that we must use a dc current that damages the junction quality leading to a decrease of the magnetoresistance ratio and therefore a smaller microwave output power compared to the case of positive currents.

\begin{figure}
   \centering
    \includegraphics[width=8.5 cm]{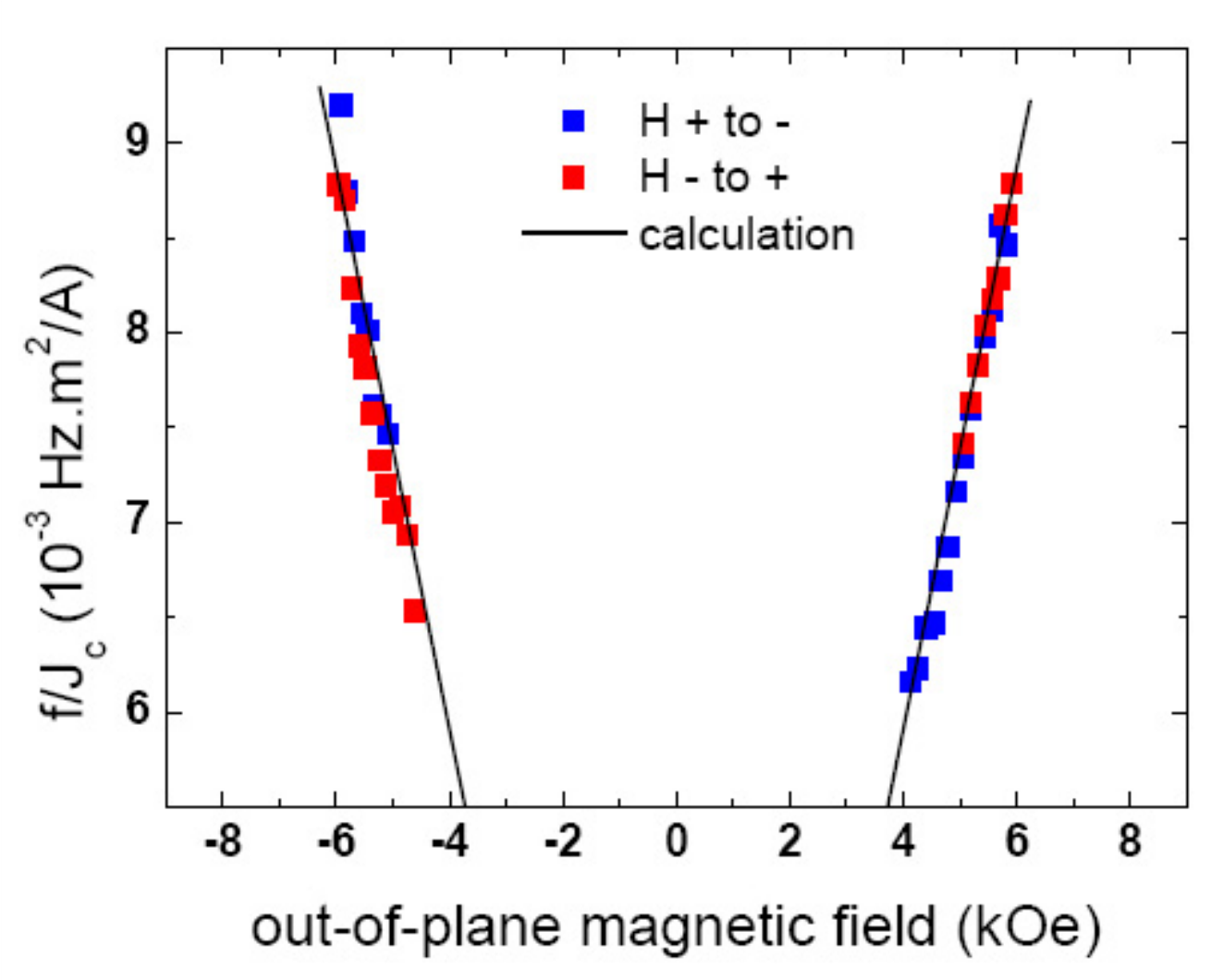}
     \caption{ratio of frequency over critical current density as a function of the out of plane magnetic field for both sweep directions (red symbols : negative to positive, blue symbols : positive to negative). The black lines correspond to the calculations using equation \ref{f/Jc}.}
\label{fig4}\end{figure}

To extract some quantitative informations about the spin torque force, an expression of the critical current density $J_c$ for the onset of sustained vortex oscillations can be obtained by equalizing Eq. \ref{STT} and Eq.\ref{damping}. By this calculation, we predict that the ratio $f/J_c$ between the frequency and the critical current density should be proportional to the out-of-plane polarization $p_z$:
\begin{equation}\label{f/Jc}
\frac{f}{J_c} =  \frac{1}{8 \pi} \left(\frac{\gamma \hbar P}{e L M^{\small{NiFe}}_{s} \alpha \eta}\right) p_z, 
\end{equation}
In our magnetic system, the spin polarization is coming from the CoFeB layer of the synthetic antiferromagnet. Thus the evolution of $p_z$ with the out-of-plane magnetic field can be expressed by : $p_z = H_{perp}/M^{\small{SAF}}_{s}$. The ratio $f/J_c$ is expected to vary linearly with $H_{perp}$. In Fig.\ref{fig4}, we plot the experimental values of $f/J_c$ vs. the out-of-plane magnetic field for increasing (blue squares) and decreasing (red squares) values of the field. The expected linear variation of $f/J_c$ with $H_{perp}$ is clearly confirmed by the experiments. In Fig.\ref{fig4}, we also plot in plain lines the result from an analytical calculation using Eq. \ref{f/Jc} taking $M^{\small{NiFe}}_{s}$ = 9.0 kOe, $M^{\small{SAF}}_{s}$ = 10.3 kOe, $\alpha$ = 0.01, $\eta$ = 1.48 and P$_{spin}$ = 0.53. We emphasize that this value of P$_{spin}$ is consistent with the MR ratios in conventional CoFeB/MgO/CoFeB MTJs in which spin polarization P$_{spin}$ = 0.5 - 0.6 for the CoFeB/MgO interface is found using the simple Julliere's model. The excellent agreement between the experimental results and the theoretical predictions on the whole field range reveals that not only a qualitative but also a quantitative understanding of the spin transfer vortex dynamics has been achieved.

\begin{figure}
   \centering
    \includegraphics[width=17 cm]{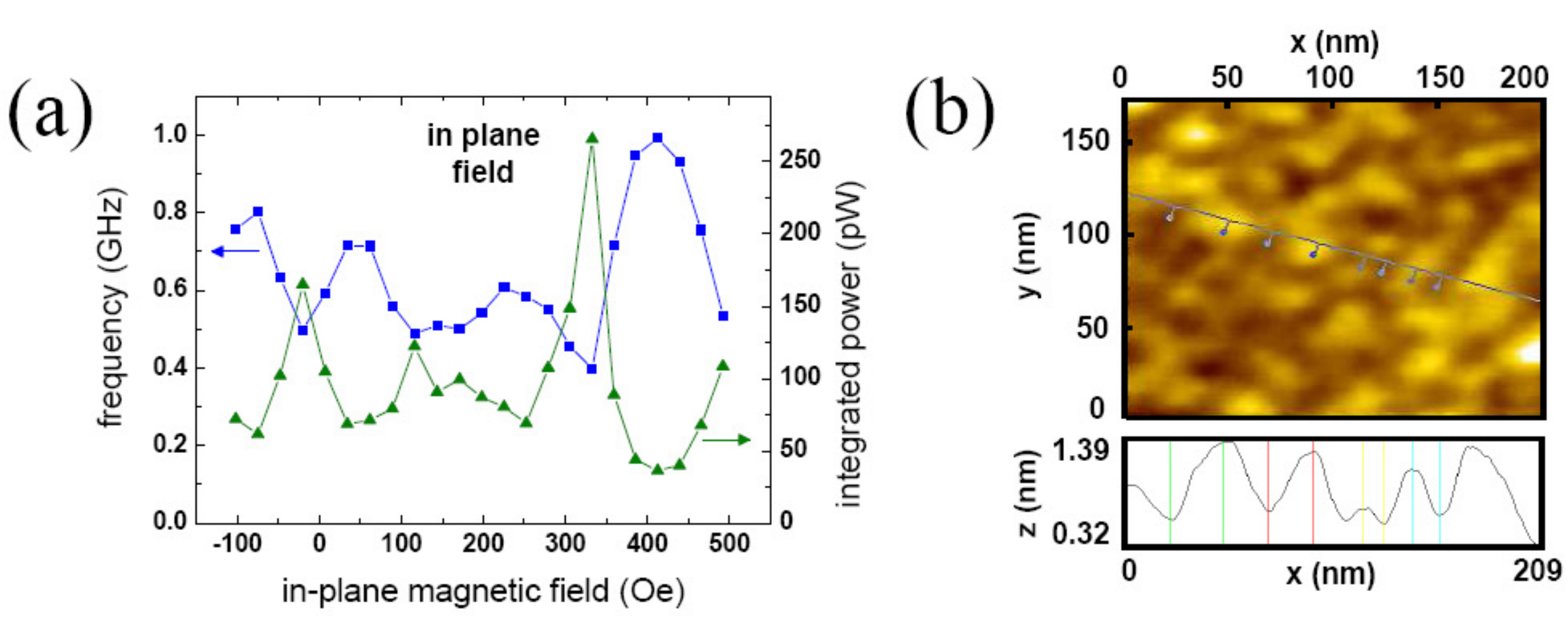}
     \caption{(a) left axis (blue squares) frequency, right axis (green triangles) : integrated power as a function of the in plane magnetic field .(b) Atomic Force Microscopy topography image of the NiFe layer. Bottom : scan of the thickness as a function of the location.}
\label{fig5}
\end{figure}

The location of the vortex core inside the nanodot can also strongly influence its dynamical properties. This core position can be controlled by the application on an in-plane magnetic field $H_{in}$ that forces the vortex to move perpendicularly to the field \cite{Guslienko:APL:2001}. In addition, it provides a tool to investigate the role of the material grains on the motion of the vortices but also potentially to increase the accessible frequencies. In Fig.\ref{fig5} (a), we display the measurement of the gyrotropic mode frequency as a function of $H_{in}$. We observe an oscillating behavior of the frequency while the magnetic vortex is traveling from the center of the dot where it nucleates, to the edge of the disk where it annihilates. Such large frequency variations are due to the modification of the magnetization and/or the local anisotropies at the grain boundaries that changes the pinning potential \cite{Compton:PRL:2006,Uhlig:PRL:2005,Vansteenkiste:NJP:2009}. From this measurement, the number of material grains in the nanodot can be evaluated. We deduce a grain size of about 40 nm in agreement with the AFM measurements shown in Fig.\ref{fig5} (b). Interestingly, we find that when the resonance frequency is low (resp. high), the thermally induced vortex fluctuations are enhanced (resp. reduced) as shown in Fig.\ref{fig5} (a); this is because the stiffness of the potential well probed by the vortex core is reduced for lower frequency. 

The perspective opened by our results is very promising. From the application point of view, we have shown that the current-induced excitation of a vortex in a low resistivity MTJ showing a MR ratio of only 10 $\%$, can already lead to microwave emissions combining large power (5 nW) and narrow linewidth (1 MHz). Furthermore the excellent fit between our experiments and the theory makes our results turn out as a textbook example of the physics of current-induced vortex gyrations. Interestingly, our work also indicates that the observation of sustained vortex gyration at zero magnetic field, required for targeted applications, is only possible either if an out-of-plane polarizer is used (as in this work by applying an out of plane field) or if the polarizer is non-uniform or dynamically excited \cite{Pufall:PRB:2007, Mistral:PRL:2008}. The quantitative agreement between our results and theory will also allow us to define the best conditions, in terms of device structure (for example tilted or out-of-plane polarizer) and materials parameters (lower R.A product and larger MR ratios), to improve the efficiency of the spin transfer and the microwave power.

\textbf{Methods} :

The magnetic stacks grown by sputtering in a CANON ANELVA chamber contain a synthetic antiferromagnet (SAF) and a free layer of NiFe separated by a thin MgO insulating barrier : PtMn 15/ CoFe 2.5 / Ru 0.85 / CoFeB 3 / MgO 1.075 / NiFe 15 /Ru 10 (nm). Details of the growth and fabrication process have been presented elsewhere \cite{Yuasa:JPhysD:2007}. The RA product is 1.3 $\Omega$.$\mu$$m^{2}$ for the parallel magnetization configuration. At room temperature, the TMR ratio is 14 $\%$ under a bias current $I_{dc}$ of 1.6 mA. The topology of the top NiFe layer has been investigated by means of Atomic Force Microscopy (see Fig.\ref{fig5} (b)). We find that the surface roughness is mainly due a spacial distribution of the thickness of the poly-crystal NiFe layer. Indeed, the surface of the underlying MgO layer was observed to be much smoother than the NiFe layer.

Prior to the investigation of transport and microwave properties, a specific study has been performed by high-resolution spin-SEM images (SEMPA) obtained at zero magnetic field to characterize the actual magnetic structures on arrays of unconnected circular MTJs with different diameters $d$ (see Additional materials). For $d$ = 72 nm, the remanent magnetic state is a single-domain state for all the nanodots. By increasing the diameter to $d$ = 125 nm, a few dots remain in single domain state whereas for most of them, the direction of the magnetic moments is changing gradually in-plane leading to a contrast varying between black and white. For larger diameters, the most favorable configuration is eventually a single vortex state, in particular for the diameter of $d$ = 170 nm chosen for this study. 

The microwave response associated to the current induced vortex dynamics is studied by applying the magnetic field $H_{perp}$ out of plane and then by sweeping the dc current $I_{dc}$ from 0 to 3.2 mA. At each current value, microwave measurements up to 1.5 GHz are recorded on a spectrum analyzer after 32 dB amplification. The background noise, measured at zero dc current, is subtracted to the power spectra. In our convention, a positive current is defined as electrons flowing from the NiFe magnetic layer to SAF layer.

For the micromagnetic simulations, we use our finite-difference micromagnetic code SPIN PM. The simulated NiFe disk has a diameter of 170 nm, with a thickness of 15 nm. The mesh cell size is set to 2 $\times$ 2 $\times$ 3.75 nm$^{3}$. We took the following magnetic parameters: $\alpha$ = 0.01 for the Gilbert damping, $M^{\small{NiFe}}_{s}$ = 9.0 kOe for the magnetization of the NiFe layer, $M^{\small{SAF}}_{s}$ = 10.3 kOe  for the effective magnetization of the polarizer (extracted from the experimental TMR curves). The spin polarization has been set to P$_{spin}$ = 0.53 (optimum polarization to reproduce the experimental frequency versus field curve of Fig.\ref{fig2}).

\newpage

\textbf{Additional figures} :

\begin{figure}[!h]
   \centering
    \includegraphics[width=8.5 cm]{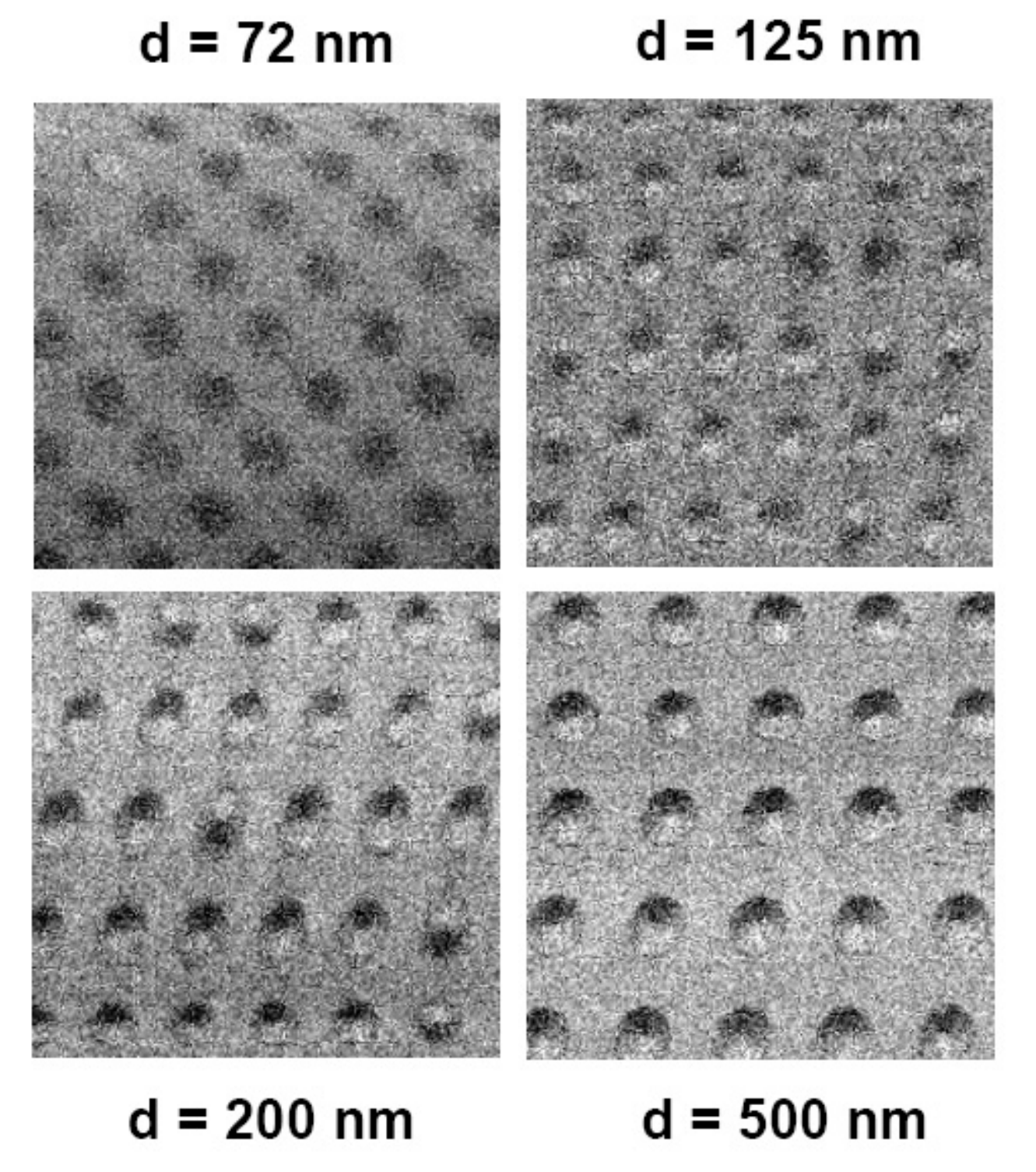}
     \caption{SEMPA images of circular 15 nm thick NiFe nanodots with different diameters d = 72, 125, 200 and 500 nm. For d = 72 nm, the remanent magnetic state is a single-domain state for all the nanodots. By increasing the diameter to d = 125 nm, a few dots remain in single domain state whereas for most of them, the direction of the magnetic moments is changing gradually in-plane leading to a changing contrast between black and white. For larger diameters, the most favorable configuration is eventually a single vortex state, in particular for the diameter of d = 170 nm chosen for the rest of the study.}
\label{AddFig1}
\end{figure}

\begin{figure}[!h]
   \centering
    \includegraphics[width=8.5 cm]{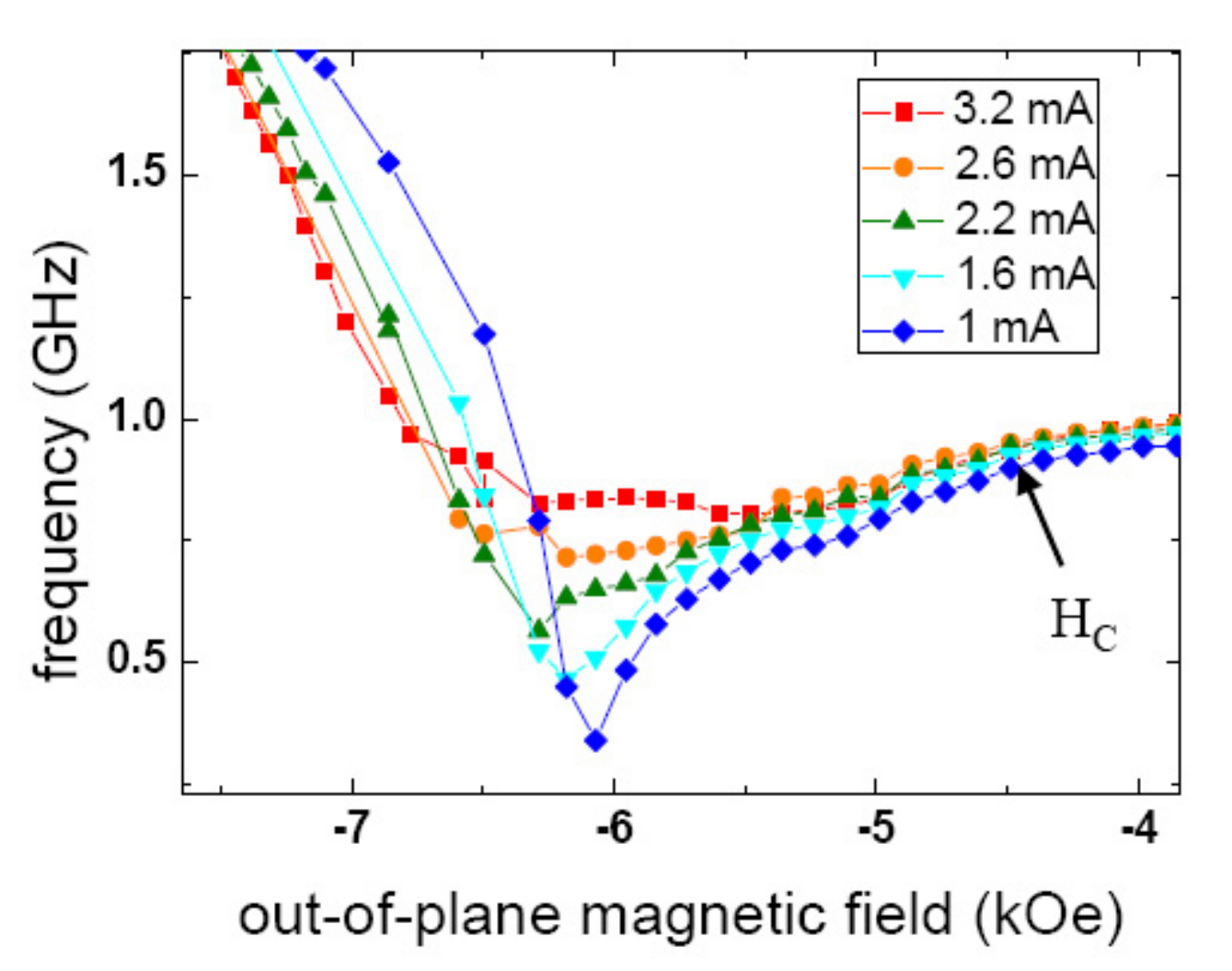}
     \caption{frequency as a function of the out-of-plane magnetic field for Idc= 1, 1.6, 2.2, 2.6, and 3.2 mA. The field is swept from negative to positive values. The combined action of the spin transfer torque and the Oersted field tends to force the magnetization of the vortex tail to remain in the plane before the vortex is annihilated and therefore restrains the Kittle mode range to higher values of the applied magnetic field.}
\label{AddFig2}
\end{figure}

\begin{figure}[!h]
   \centering
    \includegraphics[width=8.5 cm]{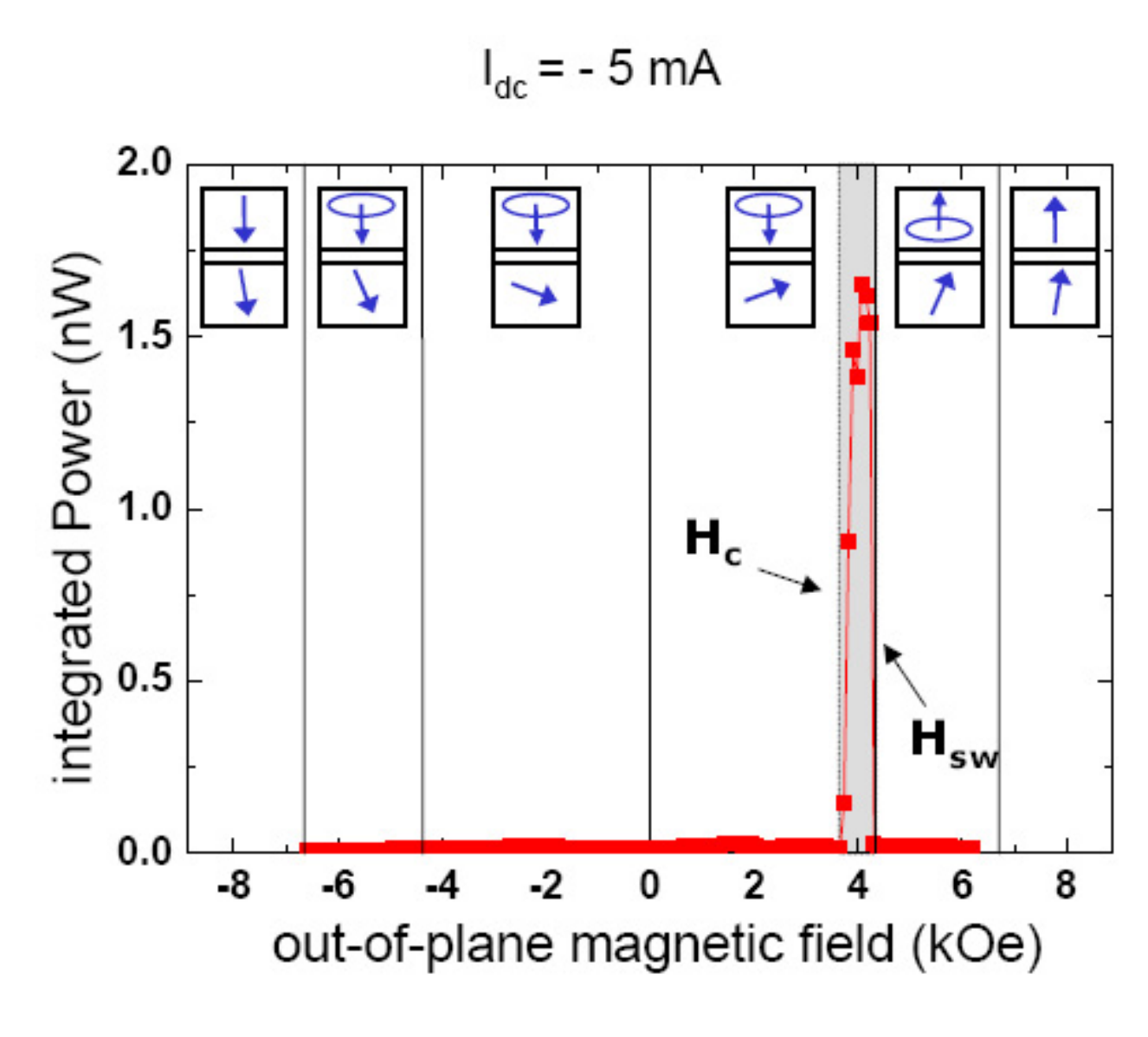}
     \caption{integrated power as a function of the out-of-plane magnetic field. The field is swept from negative to positive values. Hc is the critical field at which microwave oscillations occur for Idc = -5 mA, Hsw is the field at which the vortex core reverses.}
\label{AddFig3}
\end{figure}

\newpage

\textbf{Acknowledgments} : 

The authors acknowledge Y. Nagamine, H. Maehara and K. Tsunekawa of CANON ANELVA for preparing the MTJ films. Financial support by the CNRS and the ANR agency (NANOMASER PNANO-06-067-04, ALICANTE PNANO-06-064-03, VOICE PNANO-09-P231-36) and EU grant MASTER No. NMP-FP7-212257 is acknowledged. B. G. is supported by a PhD grant from the DGA.

\end{document}